\documentstyle[11pt]{article}
\newfont{\bbbold}{msbm10 scaled \magstep1}

\def\bbR{\mbox{\bbbold R}}

\newfont{\goth}{eufm10 scaled \magstep1}

\def\a{\alpha}
\def\b{\beta}
\def\c{\gamma}\def\C{\Gamma}
\def\d{\delta}

\def\f{\phi}

\def\k{\kappa}
\def\l{\lambda}\def\L{\Lambda}

\def\t{\tau}
\def\th{\theta}

\def\beq{\begin{equation}}\def\eeq{\end{equation}}
\def\beqa{\begin{eqnarray}}\def\eeqa{\end{eqnarray}}
\def\barr{\begin{array}}\def\earr{\end{array}}

\def\o{\omega}
\def\del{\partial}
\def\ua{\underline{\alpha}}
\def\ub{\underline{\phantom{\alpha}}\!\!\!\beta}

\def\uf{\underline\phi}
\def\una{\underline a}
\def\unb{\underline b}\def\unB{\underline B}
\def\unc{\underline c}
\def\und{\underline d}
\def\une{\underline e}
\def\unf{\underline{\phantom{e}}\!\!\!\! f}\def\unF{\underline F}
\def\ung{\underline g}
\def\unM{\underline M}

\def\unH{\underline{H}}
\def\unF{\underline{F}}\def\unT{\underline{T}}\def\unR{\underline{R}}

\def\xz{\times}

\def\nab{\nabla}\def\unab{\underline{\nabla}}

\def\nno{\nonumber}

\let\la=\label

\let\bm=\bibitem

\def\bd{\begin{document}}
\def\ed{\end{document}}
\def\ba{\begin{array}}
\def\ea{\end{array}}
\def\bea{\begin{eqnarray}}
\def\eea{\end{eqnarray}}
\def\ft#1#2{{\textstyle{{\scriptstyle #1}\over {\scriptstyle #2}}}}
\def\fft#1#2{{#1 \over #2}}
\newcommand{\be}{\begin{equation}}
\newcommand{\ee}{\end{equation}}
\newcommand{\eq}[1]{(\ref{#1})}
\def\eqs#1#2{(\ref{#1}-\ref{#2})}
\def\det{{\rm det\,}}
\def\tr{{\rm tr}}
\newcommand{\ho}[1]{$\, ^{#1}$}
\newcommand{\hoch}[1]{$\, ^{#1}$}
\newcommand{\tamphys}{\it\small Center for Theoretical Physics, 
Texas A\&M University, College Station, TX 77843, USA}
\newcommand{\newton}{\it\small Isaac Newton Institute for Mathematical Sciences,
Cambridge, UK}
\newcommand{\kings}{\it\small Department of Mathematics, King's College, London, UK}

\newcommand{\auth}{\large P.S. Howe\hoch{1}, E. Sezgin \hoch{2\dagger} 
and P.C. West \hoch{3\ddagger}}

\begin{document}

\hfill{KCL-TH-97-29}

\hfill{CTP TAMU-24/97}

\hfill{NI-97027-NQF}

\hfill{hep-th/9705093}

\hfill{\today}

\vspace{20pt}

\begin{center}
{\Large{\bf Aspects of Superembeddings}}
\vspace{30pt}

\auth

\vspace{15pt}

\begin{itemize}
\item [$^1$] \kings
\item [$^2$] \tamphys
\item [$^3$] \newton
\end{itemize}

\vspace{60pt}

{\bf Abstract}

\end{center}

Some aspects of the geometry of superembeddings and its application to
supersymmetric extended objects are discussed. In particular, the
embeddings of $(3|16)$ and $(6|16)$ dimensional superspaces into
$(11|32)$ dimensional superspace, corresponding to supermembranes and
superfivebranes in eleven dimensions, are treated in some detail. 

\vskip 5cm
\leftline{A contribution to the D.V. Volkov memorial volume}

{\vfill\leftline{}\vfill
\vskip	10pt
\footnoterule
{\footnotesize
\hoch{\dagger} Research supported in part by NSF Grant	PHY-9411543 \vskip -12pt} 
\vskip	10pt 
{\footnotesize
\hoch{\ddagger} Permanent Address: \kings \vskip -12pt}}

\pagebreak
\setcounter{page}{1}

\section{Introduction}

One of the many contributions that D.V. Volkov made to modern
theoretical physics was the realisation that supersymmetric particles,
moving in three or four dimensional spacetimes, can be described using a
formalism which has both worldline and spacetime supersymmetries built
in \cite{stv,stvz}. Up until then it had been thought that, although the
superstring can be described with either worldsheet \cite{r,ns,gso} or
spacetime supersymmetry \cite{gs}, all other extended supersymmetric
objects, including the superparticle \cite{bs}, could only be written
with manifest spacetime supersymmetry \cite{hlp,bst}. The spacetime
supersymmetric formalism does involve a local fermionic symmetry, called
$\k$-symmetry \cite{siegel}, but the geometric nature of this symmetry
remained obscure until the work of the Kharkov group showed that it can
be derived from, and is equivalent to, local worldsurface supersymmetry.
Subsequently the formalism has been developed by the Kharkov group and
others, and has been applied to various other supersymmetric extended
objects (see \cite{bstpv} for a full list of references). It also
gradually became clear that the formalism can be understood in terms of
the embedding of one superspace, the worldsurface, into another, the
target superspace. Although this point of view was implicit in the early
papers it was made explicit in a study of the heterotic string
\cite{dghs} and was further developed in \cite{bstpv}. More recently, in
\cite{hs1}, it was shown that all supersymmetric extended objects can be
understood in this way, including objects such as the Dirichlet branes
of string theory which have additional physical worldsurface bosonic
fields to the usual transverse coordinate fields. The latter are scalars
on the worldsurface whereas the new fields are gauge fields; we shall
refer to the two types of object as type I (scalars only) and type II
(additional gauge fields). The formalism was applied in particular to
construct the full equations of motion of the five-brane in
eleven-dimensional superspace \cite{hs2,hsw1}, an object that plays an
important role in $M$-theory. Moreover, it turns out that all branes are
described by the same simple embedding condition which is extremely
natural from the point of view of supergeometry \cite{hs1}. We refer the
reader to the literature for discussions of the component ($GS$)
approach to Dirichlet branes \cite{d0,d1,d2,d3,d4} and the
eleven-dimensional five-brane \cite{f1,f2,f3,f4,f5,f6,f7,f8,f9}. 

Before describing superembeddings in more detail it is perhaps
worthwhile recalling the problem that Volkov and his collaborators
solved. Consider a superparticle moving on a superworldline parametrised
by (even, odd) coordinates $(t,\t)$ in flat $(3|2)$-dimensional
superspace coordinatised by $(x^a,\th^{\a})$. Expanding the
supercoordinates describing the particle in $\t$ we have 
\beqa
x^a(t,\t) &=&x^a(t) + \t \l^a(t)\ , \nno\\
\th^{\a}(t,\t)&=&\th^{\a}(t) + \t u^{\a}(t) \ .
\la{volk}
\eeqa
The problem seems to be that there are two ``wrong statistics'' fields,
$\l^a$ and $u^{\a}$. The solution to this problem found by the Kharkov
group was to identify $u$ as a twistor variable, in fact, as the
``square-root'' of the lightlike momentum of the particle, and to regard
$\l$ as an auxiliary field. This is summarised in the superspace
equation 
\beq
Dx^a-{i\over2}D\th^{\a}(\c^a)_{\a\b}\th^{\b}=0\ ,
\la{stv}
\eeq
where $D={\del\over\del\t}+{i\over2}\t{\del\over\del t}$ is the
superworldline covariant derivative. The first component of \eq{stv} (in
a $\t$-expansion) allows one to solve for $\l$ while the second
component relates $\dot x$ to $u^2$. Volkov and his team were able to
find a somewhat unusual Lagrangian which gives rise to these conditions
on the fields $\l$ and $u$, but in what follows we shall not have much
to say about actions, rather we focus on the dynamics of the extended
objects directly and show how these can be understood from the
perspective of superembeddings. It is important to notice that the
geometrical interpretation of \eq{stv} is that, at any point on the
superworldline, the odd tangent space of the superworldline is a
subspace of the odd tangent space of the target superspace. 

\section{Flat Branes}

We define a flat brane to be an embedding of a flat superspace, of
dimension $(d|{1\over2}D')$ in a flat superspace of dimension $(D|D')$.
The existence of such objects determines in which dimensions one can
have branes and the structure of the worldsurface multiplets can be
obtained by considering small deformations. In fact, the allowed
super-dimensions correspond to the points on the modified brane scan
\cite{dl,hs1}. One could also consider branes which preserve fewer than
half of the target space supersymmetries, but we shall not do so here.
In order for flat branes to exist it is necessary that the $\C$-matrices
should decompose in the right way. If $(x^{\una},\th^{\ua}),\
\una=0,1,\dots D-1;\ \ua=1,\dots D'$ are coordinates on the target
superspace split into $(x^a,\th^{\a}),\ a=1,0,\dots d-1;\ \a=1,\dots
{1\over2}D'$ and $(x^{a'},\th^{\a'}),\ a'=d,\dots D-1;\
\a'={1\over2}D'+1,\dots D'$, we require that the $\C$-matrices split as
follows: 
\beq
(\C^{\una})_{\ua\ub}\rightarrow (\C^{a})_{\a\b},\
(\C^{a'})_{\a\b'}=(\C^{a'})_{\b'\a},\ (\C^a)_{\a'\b'}\ , 
\eeq
with all other components vanishing. If this is the case we have
\beq
[D_{\a},D_{\b}]=i(\C^a)_{\a\b} \del_a\ ,
\eeq
or, equivalently, there is a subalgebra of the supertranslational
algebra of the required dimension. The covariant derivative is defined
as usual to be 
\beq
D_{\ua}=\del_{\ua}+{i\over2}(\C^{\una})_{\ua\ub}\th^{\ub}\del_{\una}\ .
\la{cov}
\eeq
The brane itself is given by the embedding
$(x^a,\th^{\a})\mapsto(x^a,\th^{\a};0,0)$, or, equivalently, as the
solution of the equations $x^{a'}=\th^{\a'}=0$ in the target superspace.
The condition that the $\C$-matrices split as above tells us when branes
preserving half-supersymmetry can exist. 

We next consider a small deformation of the flat brane, for which the 
embedding becomes
\beq
(x,\th)\mapsto (x,\th;x'(x,\th),\th'(x,\th))\ ,
\eeq
where $x'$ and $\th'$ are small, so that we only need to work to first
order in these variables. The odd basis tangent vectors to the
submanifold, collectively denoted $E_{\a}$, are given as the image of
the (worldsurface) $D_{\a}$ under the embedding, 
\beq
E_{\a}=D_{\a}+D_{\a}\th^{\b'} D_{\b'} +(D_{\a}
X^{b'}-{i}(\C^{b'})_{\a\b'}\th^{\b'})\del_{b'}\ .
\eeq
Note that $D_{\a}$ on the target space (which occurs on the right-hand
side of the above equation) differs from $D_{\a}$ on the brane; the
former, which is the $\a$ component of \eq{cov}, includes a term
involving $\th'$ which is absent from the latter. The even basis vectors
are 
\beq
E_a=\del_a + \del_a X^{b'} + \del_a \th^{\b'} D_{\b'}\ ,
\eeq
where
\beq
X^{a'}=x^{a'} +{i\over2} \th^{\b}(\C^{a'})_{\b\c'}\th^{\c'}\ .
\eeq
We now impose the requirement that the odd tangent space at any point of
the embedded submanifold should be a subspace of the odd tangent space
of the target space at that point. This condition is required for all
supersymmetric extended objects and implies that we must impose 
\beq
D_{\a} X^{a'}={i}(\C^{a'})_{\a\b'}\th^{\b'}\ .
\la{mult}
\eeq
Computing the commutator of the odd tangent vectors (to first order in
the transverse variables) one finds 
\beq
[E_{\a},E_{\b}]=i(\C^a)_{\a\b}E_a +
i(2D_{(\a}\th^{\c'}(\C^{b'})_{\b)\c'}-(\C^a)_{\a\b}\del_a X^{b'})\del_{b'}\ . 
\eeq 
Since the commutator of two vectors of the submanifold must lead to a third we have
\beq
2D_{(\a}\th^{\c'}(\C^{b'})_{\b)\c'}=(\C^a)_{\a\b}\del_a X^{b'}\ .
\eeq
In fact, this constraint is not independent; it follows directly from
\eq{mult} by differentiation, as indeed it must since the algebra of the
covariant derivatives on the brane is preserved. 

Equation \eq{mult} above is the key equation for branes since it
determines the structure of the worldsurface supermultiplet. It can be
one of three types: on-shell, in which case it leads directly to the
dynamics of the physical fields; off-shell Lagrangian, in which case it
determines an off-shell multiplet which can be used in a Lagrangian to
determine the dynamics; or off-shell non-Lagrangian, in which case the
multiplet is off-shell but there is not a Lagrangian, at least of
conventional type, which can be constructed which leads to the dynamics.
In the third case further conditions are required, but most examples
fall into the first two classes.

In eleven dimensions (with 32 odd dimensions), there are two possible
branes, the two-brane and the five-brane\footnote{The possibility that
there might be a nine-brane has been raised \cite{hs1}; such an object,
if it exists, would have to have additional worldsurface fermion
fields.}. In both cases equation \eq{mult} defines an on-shell
supermultiplet, the $d=3, N=8$ scalar multiplet and the $d=6, N=2$
tensor multiplet, respectively. In both cases the leading components of
$X^{a'}$ and $\th^{\a'}$ can be interpreted as Goldstone fields
corresponding to the breaking of supertranslational symmetry, but in the
five-brane there is an extra component field which appears at leading
order in $D_{\a}\th^{\b'}$. In general one has 
\beq
D_{\a}\th^{\b'}=\frac{1}{2}(\C^{ab'})_{\a}{}^{\b'}\del_a X_{b'} + h_{\a}{}^{\b'}\ ,
\eeq
where
\beq 
h_{(\a}{}^{\c'}(\C^{a'})_{\b)\c'}=0\ ,
\eeq
but the latter equation only has non-trivial solutions for type II
branes. For example, for the eleven-dimensional five-brane one finds
\beq
h_{\a}{}^{\b'}={1\over6}(\C^{abc})_{\a}{}^{\b'}h_{abc}\ .
\eeq
The field $h_{abc}$ is totally antisymmetric and self-dual, and at the
linearised level is closed. Its leading component is therefore the
self-dual field strength tensor of a two-form gauge field. The quantity
$D_{\a}\th^{\b'}$, evaluated at $\th=0$, is the analogue of $u^{\a}$ in
equation \eq{volk}, at least in the linearised theory. The term
involving $\del_a X^{b'}$ generalises the momentum which arises in the
particle case, while the $h$-term is present only for type II branes. 

\section{${\bf D=11}$ Supergeometry}

In the rest of the paper we shall focus on superembeddings in eleven
dimensions. We briefly recall the salient features of eleven-dimensional
supergeometry. One has a real $(11|32)$-dimensional supermanifold $\unM$
with a choice of odd tangent bundle $\unF\subset\unT$, the full tangent
bundle, such that the associated Frobenius tensor $\uf$, defined by 
\beq
\uf(X,Y)=[X,Y]\, {\rm mod}\, \unF\ ,
\la{frob1}
\eeq
where $X$ and $Y$ are odd vector fields, is invariant under the group
$Spin(1,10)\xz \bbR^+$. This implies that there exist local bases
$(E_{\ua})$ for $\unF$ and $(E^{\una})$ for $\unB^*$, where $\unB$ is
the quotient of $\unT$ by $\unF$ and the star denotes dual, in which the
components of $\uf$ are given by 
\beq
\uf_{\ua\ub}{}^{\unc}=\langle[E_{\ua},E_{\ub}], E{}^{\unc}\rangle=i(\C^{\unc})_{\ua\ub}\ .
\la{frob2}
\eeq
This set up defines what one might call a special superconformal
structure; more generally one can allow for additional terms in $\uf$
involving two-index and five-index $\C$-matrices and we shall come back
to this possibility shortly. If \eq{frob2} holds it can be shown that it
implies the equations of motion of eleven-dimensional supergravity must
be satisfied modulo certain topological niceties which we shall ignore
here \cite{howe}. More precisely, one can show, given \eq{frob2}, that
one can find a choice of $\unB$ as a subbundle of $\unT$ and a choice of
$Spin(1,10)$ connection such that the torsion and curvature tensors in
superspace are related to those of on-shell supergravity by a super-Weyl
transformation. One can therefore make such a transformation to
eliminate the conformal factor thereby arriving at the standard geometry
\cite{cf,bh}. This geometry has structure group $Spin(1,10)$ and
therefore admits an invariant Lorentzian metric $\ung_B$ on $\unB$ and
also an invariant fermionic metric $\ung_F$ on $\unF$ whose components
in the standard basis are the components of the charge-conjugation
matrix. 

We note that when a connection is introduced it is natural to equate
$\uf$ with the dimension zero component of the torsion tensor (with a
minus sign), although it is a perfectly well-defined tensor belonging to
the space $\wedge^2\unF^*\otimes\unB$ even if a connection is not
introduced. Thus we have 
\beq
\uf_{\ua\ub}{}^{\unc}=-T_{\ua\ub}{}^{\unc}\ .
\eeq

\section{Embeddings}

We consider embeddings $M\stackrel{f}{\rightarrow}\unM$ of the
worldsurface $M$ into the target space $\unM$ which we shall take to
have dimension $(11|32)$ for definiteness, although the discussion below
is applicable more generally with appropriate modifications. It will be
assumed that $\unM$ has a superconformal structure but initially at
least we shall not suppose that the Frobenius tensor $\uf$ is invariant
under the structure group $Spin(1,10)\xz\bbR^+$. Without loss of
generality we can take it to be of the form 
\beq
\uf_{\ua\ub}{}^{\unc}=
i(\C^{\unc})_{\ua\ub}+(\C^{\unb\unc})_{\ua\ub}X_{\unb\unc}{}^{\una}+(\C^
{\unb\unc\und\une\unf})_{\ua\ub}Y_{\unb\unc\und\une\unf}{}^{\una}\ , 
\eeq
where the antisymmetric components of $X$ and $Y$ vanish, as well as
their traces. Both of these restrictions are compatible with the group
structure. We shall call such a structure a general superconformal
structure. 

There is a natural choice of odd tangent bundle $F$ on $M$ given by
\beq
F=T\cap\unF\ .
\la{embed}
\eeq
Dually, one has
\beq
B^*=T^*\cap\unB^*\ .
\la{embed'}
\eeq
The only other requirement we need to impose on the embedding is that
the metric iduced on $B^*$ (from any of the conformal class of
Lorentzian metrics on $\unB^*$) should be Lorentzian with signature
$(p-1),\ p=2,5$. Again this condition is automatically conformally
invariant. 

The Frobenius tensor $\f$ of $M$ is defined in the same way as above,
namely as the commutator of two odd vector fields modulo the odd tangent
bundle. The relation between the worldsurface and target space Frobenius
tensors is given by 
\beq
\uf(X,Y,\underline{\o})=\f(X,Y,f^*\underline{\o})\ ,
\la{frob3}
\eeq
where $X$ and $Y$ are odd vector fields on $M$ (which may be considered
as vector fields on $\unM$) and $\underline{\o}$ is a one-form on $\unM$. 

For any embedding one has three natural bundles (on $M$), the tangent
bundle $T$, the tangent bundle, $\unT$, of $\unM$ restricted to $M$ and
the normal bundle $T'$ which fit together in a short exact sequence 
\beq
0\rightarrow T\rightarrow \unT\rightarrow T'\rightarrow 0\ .
\eeq
However, in the super case we have even and odd tangent bundles which
themselves fit into an exact sequence 
\beq
0\rightarrow F\rightarrow T\rightarrow B\rightarrow 0\ ,
\eeq
and similarly for the target space as well as the corresponding normal
bundles. In fact there are nine bundles in all and it can be shown that
they fit together into the following diagram, 
\beq
\barr{ccccccccc}
&&0&&0&&0&&\\
&&\uparrow&&\uparrow&&\uparrow&& \\
0&\rightarrow & F'&\rightarrow& T'&\rightarrow & B'&\rightarrow& 0 \\
&&\uparrow&&\uparrow&&\uparrow&& \\
0&\rightarrow & {\unF}&\rightarrow& {\unT}&\rightarrow & {\unB}&\rightarrow& 0 \\
&&\uparrow&&\uparrow&&\uparrow&& \\
0&\rightarrow & F&\rightarrow& T&\rightarrow & B&\rightarrow& 0 \\
&&\uparrow&&\uparrow&&\uparrow&& \\
&&0&&0&&0&&
\earr
\eeq
where each of the rows and columns is exact and where each square is
commutative. The proof of these assertions is straightforward. The dual
bundles give rise to a similar diagram with the arrows reversed. In
practice one wishes to split the sequences, so that the central bundle
of each sequence becomes a direct sum of the other two. However, when
this is carried out it is important to note that, although $B$ as a
quotient bundle is a subbundle of $\unB$ the same is not true when $B$
and $\unB$ are regarded as subbundles of $T$ and $\unT$ respectively. 

\section{Brane Integrability}

The basic embedding condition described above is extremely natural given
the geometrical structures that arise in supergeometry. Moreover, it is
also extremely restrictive. In fact, in the eleven-dimensional examples
we are discussing, one has the following results: if an
$(11|32)$-dimensional supermanifold $\unM$ with a general superconformal
structure admits embeddings of the type described in the previous
section of either two-branes or five-branes through every point, then:
\begin{itemize}
\item{any such brane is dynamical, that is the embedding implies that
the worldsurface multiplet is on-shell,} 
\item{the superconformal structure on the target space must be special,
which implies, as we have discussed above, that the equations of motion
of eleven-dimensional supergravity must be satisfied,}
\item{the worldsurface supergeometry is completely specified up to gauge freedoms.}
\end{itemize}

In some respects this may not be too surprising as it has been known for
some time that the requirement of $\k$-symmetry in the $GS$ formulation
of the membrane forces the target space supergeometry to be equivalent
to eleven-dimensional supergravity \cite{bst}. However, in the present
approach we have achieved this with the bare minimum of assumptions;
everything, including $\k$-symmetry follows from the simple embedding
condition \eq{embed}.

The complete proof of the above assertions is extremely long and rather
complicated in terms of details, but it is simple to understand how it
comes about in principle. Consider the relation \eq{frob3} between the
Frobenius tensors of the two manifolds. We introduce local bases
$(E_{\a}),\ (E^a)$ for $F$ and $B^*$ respectively and note that
\eq{embed} implies that 
\beq
E_{\a}=E_{\a}{}^{\ua}E_{\ua}\ ,
\eeq
for some $16\xz 32$ matrix $E_{\a}{}^{\ua}$, while the dual condition \eq{embed'} implies
\beq
f^*E^{\una}=E^a E_{a}{}^{\una}\ .
\eeq
Equation \eq{frob3} then reads in components with respect to these bases,
\beq
E_{\a}{}^{\ua}E_{\b}{}^{\ub}\uf_{\ua\ub}{}^{\unc}=\f_{\a\b}{}^c E_c{}^{\unc}\ .
\la{frob4}
\eeq
If $(E_{\ua})$ is a spin basis for $\unF$ any other such basis will be
related to it by an element $u$ of $Spin(1,10)$ up to a conformal factor
which we shall ignore for the moment. We write
$u=(u_{\a}{}^{\ua},u_{\a'}{}^{\ua})$, with $\a'=1,\ldots 16$. Since
$E_{\a}{}^{\ua}$ has maximal rank there will be a choice of $u$ such
that $E_{\a}$ is related to $u_{\a}{}^{\ua}E_{\ua}$ by a non-singular
matrix. Hence, without loss of generality, we can write 
\beq
E_{\a}{}^{\ua}=A_{\a}{}^{\b}u_{\b}{}^{\ua}+B_{\a}{}^{\b'}u_{\b'}{}^{\ua}\ ,
\eeq
where $\det A\neq 0$. Making a change of basis for $F$ we arrive at
\beq
E_{\a}{}^{\ua}=u_{\a}{}^{\ua}+h_{\a}{}^{\b'}u_{\b'}{}^{\ua}\ .
\eeq
On the bosonic space $B^*$ the situation resembles more closely the case
of a Lorentzian embedding and we may choose, again up to a conformal
factor 
\beq
E_a{}^{\una}=u_a{}^{\una}\ ,
\eeq
where $(u_a{}^{\una},u_{a'}{}^{\una})$ is the element of $SO(1,10)$
corresponding to $u=(u_{\a}{}^{\ua},u_{\a'}{}^{\ua})\in Spin(1,10)$
Thus, at any point $p\in M$, the embedding is specified by
$u_a{}^{\una},u_{\a}{}^{\ua}$ and $h_{\a}{}^{\b'}$. 

We can now decompose equation \eq{frob4} into components tangent and
normal to $M$. We then find 
\beq
\uf_{\a\b}{}^{c'}+2h_{(\a}{}^{\c'}\uf_{\b)\c'}{}^{c'}+
h_{\a}{}^{\c'}h_{\b}{}^{\d'}\uf_{\c'\d'}{}^{c'}=0\ , 
\la{frob5}
\eeq
and
\beq
\uf_{\a\b}{}^{c}+2h_{(\a}{}^{\c'}\uf_{\b)\c'}{}^{c}+
h_{\a}{}^{\c'}h_{\b}{}^{\d'}\uf_{\c'\d'}{}^{c}=\f_{\a\b}{}^c\ ,
\la{frob6}
\eeq
where
\beq
\uf_{\a\b}{}^{c'}=u_{\a}{}^{\ua} u_{\b}{}^{\ub}\uf_{\ua\ub}{}^{\unc}u_{\unc}{}^{c'}\ ,
\eeq
and similarly for the other projections of $\uf_{\ua\ub}{}^{\unc}$. Now
in order for there to be embeddings of branes in general we require that
these equations be satisfied for arbitrary embeddings passing through a
given point $p\in \unM$ and furthermore that this should be true for all
points of $\unM$. Since we may vary $u$ and $h$ independently this
requires 
\beq
\uf_{\a\b}{}^{c'}=0\ .
\eeq  
This can only be satisfied for arbitrary embeddings if the
superconformal structure on $\unM$ is special, i.e. if 
\beq
\uf_{\ua\ub}{}^{\unc}=i(\C^{\unc})_{\ua\ub}\ .
\eeq
Given this one finds that equations \eq{frob5} and \eq{frob6} are solved by
\beq
h_{\a}{}^{\b'}=\cases{0 &two-brane \cr
{1\over6}(\C^{abc})_{\a}{}^{\b'}h_{abc} &five-brane\cr}
\eeq
where $h_{abc}$ is self-dual, and
\beq
\f_{\a\b}{}^c=\cases{i(\C^c)_{\a\b} &two-brane\cr
i(\C^b)_{\a\b} m_b{}^c &five-brane\cr}
\eeq
where
\beq
m_a{}^b=\d_a{}^b-2h_{acd}h^{bcd}\ .
\eeq

The above argument establishes brane-integrablity; to see that the
embedding implies the dynamics it is sufficient to consider the
linearised case, i.e. take the target space to be flat and assume that
the embedded submanifold is also nearly flat. It is not too difficult to
see that in this limit one recovers the equations describing the
deformations of flat branes, and hence the worldsurface fields are
indeed on-shell. By working to second order in the transverse fields one
can quickly see that the worldsurface supergravity fields are also
determined.

\section{Some Geometrical Aspects of Superembeddings}

We briefly recall some aspects of Riemannian embeddings (see, for
example, \cite{kn}). Let $M$ be a manifold embedded in a Riemannian
manifold $(\unM,\ung)$. The metric on the target space induces natural
metrics on the embedded space and on the normal bundle $T'$ as well as
determining a natural orthogonal decomposition of $\unT$ into tangential
and normal components. Explicitly 
\beqa 
\ung(X,Y) &=& g(X,Y)\ , \nno\\
\ung(X,Y') &=& 0\ , \nno\\
\ung(X',Y') &=& g'(X',Y')\ ,
\eeqa
where $X,Y$ are tangential vector fields, $X',Y'$ normal vector fields,
$g$ is the induced metric on $M$ and $g'$ is the metric induced on the
normal bundle. Metric connections $\nab$ and $\nab'$ are determined in
$T$ and $T'$ respectively from the metric connection $\unab$ on $\unM$
by the Gauss-Weingarten equations 
\beqa
\unab_X Y&=&\nab_X Y + K'(X,Y)\ , \nno\\
\unab_X Y'&=&\nab'_X Y' + K(X,Y')\ ,
\la{gw}
\eeqa
where $K'(X,Y)$ is normal and $K(X,Y')$ tangential. $K'$ is the second
fundamental form, and $K$ is related to $K'$ by 
\beq
\ung(K'(X,Y),Z')+ \ung(X,K(Y,Z'))=0\ .
\eeq
From \eq{gw} one can derive the torsion equations 
\beqa
{[\unT(X,Y)]}^t &=&T(X,Y)\ , \nno\\
{[\unT(X,Y)]}^n &=& K'(X,Y) - K'(Y,X)\ ,
\eeqa
where the superscripts $t$ and $n$ denote tangential and normal
respectively. Finally, we have the equations of Gauss and Codazzi
relating the curvature tensors of $T$ and $T'$ to the Riemann curvature
tensor of $\unM$: 
\beqa
\unR(X,Y,Z,\o)&=&R(X,Y,Z,\o) + (K(X,K'(Y,Z),\o) -X\leftrightarrow Y)\ , \nno\\
\unR(X,Y,Z',\o')&=&R'(X,Y,Z',\o') + (K'(X,K(Y,Z'),\o') -X\leftrightarrow Y)\ ,
\la{gc}
\eeqa
where $\o$ and $\o'$ are respectively tangential and normal one-forms.

The above equations can be generalised to the supersymmetric case
although the situation is more complicated due to the even-odd split. In
view of the discussion of the preceding section we can assume that the
target space supergeometry is the standard geometry describing on-shell
supergravity. We begin with the membrane. The tensor $\uf$ on $\unM$
gives rise to the following tensors via embedding: 
\beqa
\uf(X,Y,\o)&=&\f (X,Y,\o)\ , \nno\\
\uf(X,Y',\o')&=&\tilde\f (X,Y',\o')\ , \nno\\
\uf(X',Y',\o)&=&\f' (X',Y',\o)\ , 
\eeqa
while
\beqa
\uf(X,Y,\o')&=&0\ ,\phantom{\uf(X,Y,\o)} \nno \\
\uf(X,Y',\o)&=&0\ , \nno\\
\uf(X',Y',\o')&=&0\ ,
\eeqa
where, in both equations, the vectors are all odd, the forms are even
and normal vectors or forms are distinguished by a prime. From the
bosonic metric we derive 
\beqa
\ung_B(X,Y)&=&g_B(X,Y)\ , \nno\\
\ung_B(X,Y')&=&0\ , \nno\\
\ung_B(X',Y')&=&g_B'(X,Y)\ , 
\eeqa
for even tangential and normal vectors $X,Y$ and $X',Y'$, thus defining
induced metrics for $B$ and $B'$. Starting from the fermionic metric we
get 
\beqa
\ung_F(X,Y)&=&g_F(X,Y)\ , \nno\\
\ung_F(X,Y')&=&0\ , \nno\\
\ung_F(X',Y')&=&g_F'(X,Y)\ ,
\la{fermi} 
\eeqa
for odd tangential and normal vectors $X,Y$ and $X',Y'$, thus defining
induced fermionic metrics for $F$ and $F'$. We also have 
\beqa
\ung_F(X,Y)&=&0\ , \nno\\
\ung_F(X,Y')&=&0\ , \nno\\
\ung_F(X',Y)&=&\L(X',Y)\ , \nno\\
\ung_F(X',Y)&=&0\ , 
\eeqa
for odd vectors $X,X'$ and even vectors $Y,Y'$, with the primes denoting
normal vectors as usual. The above equations determine a decomposition
of $\unT$ with respect to the tangential and normal bundles. Explicitly,
we have 
\beq
\unF\cong F\oplus F'\ ,
\eeq
while
\beq
\unB\subset B\oplus B' \oplus F'\ .
\eeq
The field $\L$ can be thought of as providing a gauge-invariant
representation of the worldsurface multiplet. Indeed, in the linearised
case it reduces to the even derivative of the transverse odd coordinate
functions. 

The generalisations of the Gauss-Weingarten equations are
\beqa
\unab_X Y&=&\nab_X Y + K'(X,Y) + L(X,Y)\ ,\nno\\
\unab_X Y'&=&\nab'_X Y' + K(X,Y') + L'(X,Y')\ ,
\eeqa
where $K'(X,Y)$ and $L'(X,Y)$ are normal while $K(X,Y')$ and $L(X,Y)$
are tangential. The additional tangential terms are required because
even vectors on $M$ have non-vanishing projections on $\unF$. For $Y$
$Y'$ odd $L(X,Y)$ and $L'(X,Y')$ both vanish while $K(X,Y)$ and
$K'(X,Y')$ are odd. The torsion equations are 
\beqa
[\unT(X,Y)]^t &=& T(X,Y) + L(X,Y)-L(Y,X)\ , \nno\\
{[\unT(X,Y)]}^n &=& K(X,Y)-K(Y,X)\ ,
\eeqa
while the Gauss-Codazzi equations have the same form as in the bosonic
case despite the presence of the additional terms in the
Gauss-Weingarten equations, 
\beqa 
\unR(X,Y,Z,\o)&=&R(X,Y,Z,\o) +
(K'(X,K(Y,Z),\o) -X\leftrightarrow Y)\ , \\
\unR(X,Y,Z',\o')&=&R'(X,Y,Z',\o') + (K(X,K'(Y,Z'),\o') -X\leftrightarrow Y)\ ,
\eeqa
where the last two arguments of the curvatures are either both even or both odd.

One can obtain many relations for the tensors defined above by
differentiating the invariant tensors. It is straightforward to check
that the connections defined on $T$ and $T'$ preserve the induced
bosonic and fermionic metrics, and that the tensors constructed from
$\uf$ are also invariant. The structure groups for $F$ and $F'$ are both
$Spin(1,2)\cdot Spin(8)$, although different representations are
involved, while the structure groups for $B$ and $B'$ are $SO_o(1,2)$
and $SO(8)$, where the superscript ``$o$'' denotes the component
connected to the identity.

Many of the above equations can be taken over in the case of the
five-brane, but there are some differences. The equations for the
bosonic metric remain the same but the fermionic ones change. One finds,
instead of \eq{fermi}, the equations 
\beqa
\ung_F(X,Y)&=&h(X,Y)\ , \\
\ung_F(X,Y')&=&g_F(X,Y')\ , \\
\ung_F(X',Y')&=&0\ ,
\eeqa
for odd arguments. Note that in decomposing the eleven-dimensional
charge conjugation matrix ($\ung_F$) into $6+5$ one does not arrive at
fermionic metrics on the tangential and normal subspaces but rather at
an off-diagonal tensor which we have called $g_F$ above although it is
not a metric but rather determines an isomorphism between $F^*$ and
$F'$. The tensor $h$ departs from this expected behaviour and is a
signal of a type II brane. In index notation, 
\beq
h_{\a\b}=\frac{1}{3}(\C^{abc})_{\a\b} h_{abc}\ .
\eeq
For mixed arguments ($X's$ odd $Y$'s even) one has
\beqa
\ung_F(X,Y)&=&\L(X,Y)\ , \nno\\
\ung_F(X,Y')&=& 0\ ,\nno\\
\ung_F(X',Y)&=&0\ ,\nno\\
\ung_F(X',Y)&=&0\ .
\eeqa
Again the field $\L$ is related to the worldsurface multiplet. For the
Frobenius tensor one finds, as before, 
\beqa
\uf(X,Y,\o)&=&\f (X,Y,\o)\ , \nno\\
\uf(X,Y',\o')&=&\tilde\f (X,Y',\o')\ , \nno\\
\uf(X',Y',\o)&=&\f' (X',Y',\o)\ , 
\eeqa
and
\beqa
\uf(X,Y,\o')&=&0\ ,\phantom{\uf(X,Y,\o)}\nno\\
\uf(X',Y',\o')&=&0\ ,
\eeqa
where the vectors are odd and the forms even. However, one now has
\beq
\uf(X,Y',\o)\neq 0\ .
\eeq
In fact, this tensor is also linearly proportional to $h$.

One can take over the Gauss-Weingarten, torsion and curvature equations
formally without change, but there are differences between the two and
five-brane cases. For the five-brane the induced connections for the
even tangent and normal bundles correspond to the groups $SO_o(1,5)$ and
$SO(5)$ respectively, but the connections for the odd tangent bundles do
not give $Spin(1,5)\cdot Spin(5)$ connections. This is again due to the
type II embedding structure. One would have had this result if $h$ had
been zero, but the intervention of this term complicates matters
somewhat. An alternative procedure is to define connections which do
preserve the natural groups in both the even and odd sectors, and this
is the route that has been taken in the literature \cite{hs2,hsw1}.

We conclude with a few remarks on Wess-Zumino forms. So far we have made
no mention of these, even though they play such a crucial r\^{o}le in
the $GS$ formalism. The reason for this is that supersymmetry implies
that they are present, so that one does not have to introduce them
separately by hand in the superspace formalism. In eleven dimensional
superspace with the standard constraints it is easy to show that there
exists a closed four-form $\unH_4$ which has non-trivial components only
at dimension zero and one, the dimension one component reflecting the
presence of a non-trivial spacetime three-form potential. The pull-back
of this form defines a four-form on $M$, obviously closed, and which is
flat in the case of the membrane, i.e. its only non-vanishing component
in a standard basis is a $\C$-matrix contribution at dimension zero, and
which obeys the equation 
\beq
dH_3=-\frac{1}{4}f^*\unH_4\ ,
\la{wz}
\eeq
in the case of the five-brane. The only non-vanishing component of $H_3$
is the purely vectorial component which is given by 
\beq
H_{abc}=(m^{-1})_a{}^d h_{bcd}\ .
\eeq
We emphasize that \eq{wz} is not a new equation; it is identically true
provided that one defines $H_3$ as above and uses the results which
follow from the torsion equations of the embedding. We refer the reader
to the literature for more details on how one deduces the full equations
of motion describing the brane dynamics from the basic embedding
condition \eq{embed} using the superspace formalism
\cite{hs2,hsw1,hsw2,hsw3}.

\pagebreak
 
\end{document}